\documentclass[sigconf,nonacm]{acmart}
\pdfmapfile{+libertine.map}
\microtypesetup{expansion=false}
\setkeys{acmart.cls}{balance=false}

\setcopyright{none}
\usepackage{amsmath}
\usepackage{algorithm}
\usepackage{algorithmic}
\usepackage{balance}
\usepackage{dblfloatfix}
\usepackage{needspace}
\usepackage{placeins}
\usepackage{siunitx}
\newcommand{\pNinety}{Q_{0.9}}
\newcommand{\cellgap}{D_{\mathrm{cell}}}
\newcommand{\caprate}{\alpha_L}
\newcommand{\tablelayout}{%
  \renewcommand{\arraystretch}{1.04}%
  \setlength{\tabcolsep}{3.0pt}%
}

\title[Audits with Common Evidence]{Freeze, Validate, Report: Auditing Urban Station Plans with Common Evidence}

\author{Julian Teusch}
\email{julian.teusch@tu-clausthal.de}
\affiliation{%
  \department{Institute of Computer Science}
  \institution{Clausthal University of Technology}
  \city{Clausthal-Zellerfeld}
  \country{Germany}
}
\author{Oliver Kesz{\"o}cze}
\email{oliver.keszoecze@tu-clausthal.de}
\affiliation{%
  \department{Computer Engineering}
  \institution{Clausthal University of Technology}
  \city{Clausthal-Zellerfeld}
  \country{Germany}
}
\renewcommand{\shortauthors}{Teusch and Kesz{\"o}cze}

\begin{document}
\raggedbottom

\begin{abstract}
Transport authorities comparing urban station plans need to know whether score
differences reflect the plans or their evaluation inputs. If spatial units,
included trips, or candidate blocks are reconstructed from each plan's access
outcomes, apparently comparable scores can refer to different evidence bases.
\emph{Freeze--Validate--Report} fixes an evaluation contract before comparison:
target trips, proxy groups, reporting cells, pools of candidate blocks, and a
predeclared lexicographic selection rule. A two-point example shows that
plan-specific observation selection can reverse the target mean ordering.
We compare seven station-plan generators spanning classical clustering and
facility location, fairness-oriented adaptations, and a grid control. We score
every plan on common validation evidence using endpoint distance, the sum of the
distances from a trip's origin and destination to their respective nearest
stations; a five-metric rule selects one for a single
test report.
For Porto, the rule selects IFkCO, our adaptation of individually fair
$k$-center with outliers; its test overall and worst-group 90th percentiles (p90)
are \qty{805}{\meter} and \qty{872}{\meter}. For Chicago, the rule selects Grid,
a grid-based control; the corresponding values in a later test window on the
same day are \qty{1404}{\meter} and \qty{1656}{\meter}. Two post hoc checks show
why the contract records its inputs: expanding the candidate set from 300 to 600
changes the selection in both cities, while a Chicago design that holds the hour
fixed across three dates selects Priority, a priority-center adaptation. A secondary proportional
mean fairlet (PMF) diagnostic shows that plan-specific proposal pools can change
diagnostic ranks; admissibility and solver bounds apply only to the sampled pool.
The package supplies code, hashes, decision logs, and proposal pools. We establish
auditability within a declared contract, not stable performance after deployment;
proxy definitions and metric priorities remain the authority's choices.
\end{abstract}



\maketitle
\begingroup
\renewcommand{\thefootnote}{}
\footnotetext[0]{Author preprint. Accepted for publication in the proceedings of the 4th ACM SIGSPATIAL International Workshop on Advances in Urban-AI (UrbanAI 2026).}
\endgroup

\section{Introduction}
\label{sec:introduction}

Consider a transport authority choosing among proposals for $k$ mobility
stations. A station plan is a set of $k$ candidate sites. The authority wants
to compare overall access and access across area-level proxy groups. For an
observed trip, endpoint distance sums the origin-to-nearest-station and
destination-to-nearest-station distances. The comparison is meaningful only when
every plan is scored on the same trips and reporting cells under the same selection rule. If
those inputs are reconstructed from each plan's own outcomes, the authority
cannot tell whether a favorable score reflects better access or a more favorable
evidence base.
Section~\ref{sec:common-evidence} gives an example with two demand points in
which such
reconstruction reverses the ranking.

Station placement distributes access burdens across neighborhoods. Classical
facility location objectives summarize this burden through a demand-weighted
total or a maximum distance
\cite{gonzalez1985kcenter,yu2009transithub}. Other methods
instead optimize the largest burden among groups or a priority objective
\cite{jung2020service,bajpai2021priority,ghadiri2021socially}. A planning audit
may also group trips into spatial reporting units and ask whether
fixed proxy groups receive comparable outcomes.

Exact optimization does not resolve this evaluation problem: an exact method
only optimizes the supplied instance. Without a separately proved invariance
condition, comparable scores require a common instance. The post hoc checks in
Section~\ref{sec:sensitivity-results} illustrate sensitivity to these inputs: changing the
candidate set or temporal window changes the selected plan. The contract records
this dependence; whether its fairness definition is appropriate remains a
separate normative question.

Throughout, the planning authority declares the comparison and chooses; the
analyst applies the declared inputs; and the auditor checks the record. The
\emph{evaluation contract} comprises all inputs declared before
a comparison. \emph{Common evidence} denotes the observations, reporting
partition, and pool of candidate blocks shared across plans. The \emph{selection rule}
denotes the ordered validation metrics used to choose a plan.

We study the following question:

\begin{quote}
\textbf{Research question.} Can a planning authority compare station plans on a
common evidence base while keeping evidence construction, validation-based
selection, and final test reporting traceable and separate?
\end{quote}

\emph{Freeze--Validate--Report} has three phases. First, the authority records its
choices; training data determine candidate sites, plans, and the coordinate
frame.
Second, the analyst scores all plans on common validation evidence, and the
authority's rule chooses one; the Pareto front is descriptive only. Third, the
program hashes the selected plan and decision record before loading test data,
then evaluates only that plan. Test outcomes cannot alter the choice.

We evaluate two public origin--destination (OD) settings. Porto taxi trips use
four census socioeconomic status (SES) strata (Porto-SES). Chicago ride-hail
trips use four Hardship Index strata defined from community areas
(Chicago-Hardship).
In both settings, we treat observed trip endpoints as demand for hypothetical
stations: the datasets contain observed OD trips, not trips made between the
hypothetical stations.
Section~\ref{sec:setup} gives the sources, periods, and sampling procedure.
The seven generators span classical clustering and facility location,
fairness-oriented adaptations, and a grid control. Together, they let us study
comparative evaluation for urban decision intelligence while holding the
decision target fixed. Learned or hybrid generators can enter as additional
plan sources under the same contract.

The contributions are:

\begin{itemize}
  \item a three-phase evaluation protocol that freezes the comparison contract,
  selects one plan on validation data, and reports it once on test data;
  \item an audit-consistency condition that identifies when plan scores refer
  to the same observations and comparison inputs, together with a minimal
  rank-reversal example;
  \item a two-city comparative evaluation of seven station-plan generators
  spanning clustering, facility location, fairness-oriented adaptations, and a
  grid control, using every sampled trip, a grid and validity thresholds fixed
  before validation, and a predeclared lexicographic selection rule; and
  \item a secondary proportional mean fairlet (PMF) ablation of common versus plan-specific proposal pools,
  post hoc design and bootstrap sensitivity checks, and a reproducibility package with phase logs,
  hashes, proposal pools, and separately implemented verification code.
\end{itemize}

These retrospective demonstrations were neither used nor validated by the Porto
or Chicago authorities and do not estimate how performance or rankings vary
after deployment.

\section{Related Work and Positioning}
\label{sec:related}

Urban AI systems produce predictions, spatial representations, and
infrastructure analyses that can inform planning decisions
\cite{pranto2025hubs,namgung2025carewell,lou2025urbanmas,scherrer2025infrastructure}.
We study the comparison layer of urban decision intelligence: when classical,
learned, or hybrid methods supply alternative plans, a planning authority needs
scores that refer to a common target and common evidence.

Model cards report intended uses, performance across groups, and evaluation
procedures for trained models \cite{mitchell2019modelcards}. End-to-end internal
audit frameworks call for traceable records across development stages
\cite{raji2020accountability}. Our condition is narrower: it formalizes when
alternative plans are scored against the same observations and evaluation
inputs.

Classical $k$-center and $p$-median objectives motivate our geometric and
demand-weighted baselines \cite{gonzalez1985kcenter,yu2009transithub}. Jung et al.\ define
individual service relative to a local neighborhood radius
\cite{jung2020service}; priority-center work likewise scales service distances
by a radius defined for each point and accommodates outliers
\cite{bajpai2021priority}.
Socially fair clustering controls the largest group cost
\cite{ghadiri2021socially}, while individually fair $k$-center with outliers
(IFkCO) combines a local service requirement with explicit exclusions
\cite{han2023ifkco}. We adapt these ideas to choose stations from a finite set of
sites. We compare the resulting plans, not the original algorithm
families. The audit includes every validation and test trip even when a training objective
allows outliers.

Fairlets are small blocks with a prescribed group composition that can then be
clustered as indivisible units \cite{chierichetti2017fairlets}; fair
clustering with outliers motivates explicit accounting for who is left out
\cite{almanza2022fairoutliers}. Our secondary blocks impose exact counts for each
proxy group, a physical radius, and a limit on differences in mean station access
among groups. We use these blocks only for diagnosis. Hierarchy construction and fairlet
approximation guarantees lie outside the study.

Spatial equity results can also change with the scale and zoning of reporting
units. This dependence is known as the modifiable areal unit problem (MAUP).
Javanmard et al.\ demonstrate such sensitivity for public-transit reliability equity
\cite{javanmard2023maup}. We fix one grid to prevent each plan from choosing its
own units; the choice of scale is subject to MAUP.

Work on adaptive data analysis separates exploratory reuse from valid holdout
inference \cite{dwork2015reusable}. Our narrower concern is that an analyst must
first score two plans against the same stated target. Roberts et al.\
recommend blocked validation when observations have temporal or spatial
dependence \cite{roberts2017crossvalidation}.
Our primary splits preserve temporal order but use only one sequence; blocked or
rolling-origin validation would require additional windows.

Work on measurement models emphasizes that an operational proxy may differ from
its intended construct \cite{jacobs2021measurement}. Likewise, transport
accessibility includes land use, opportunities, time, and individual constraints
beyond a distance summary \cite{geurs2004accessibility}. These partial measures
must therefore be declared as decision inputs rather than treated as neutral
descriptions of a city.

\section{Problem and Evaluation Contract}
\label{sec:problem}

We use $\mathcal D$ for an evaluation dataset
and $A$ for an audit implementation.

\subsection{Common evidence and rank reversal}
\label{sec:common-evidence}

An earlier PMF constructor in our own pipeline used each plan's origin-access
outcomes to assign trips to block anchors, the seed points from which blocks
are grown. It therefore supplied a different
proposal pool to each plan's diagnostic optimization, even with trips and
constraints held fixed. Section~\ref{sec:ablation-setup} describes this concrete
instance of plan-dependent evidence; the following simplified example isolates
the effect of changing the observations being scored.

\paragraph{Example with two demand points.}
Suppose a planning authority declares mean access burden across two
neighborhoods, represented by one demand point each. Plan $S_1$ yields
\qty{0}{\meter} and \qty{1000}{\meter}, with mean \qty{500}{\meter}; plan $S_2$
yields \qty{400}{\meter} and \qty{500}{\meter}, with mean \qty{450}{\meter}.
The declared target therefore ranks $S_2$ first. An analyst instead retains each
plan's best-served point and reports \qty{0}{\meter} for $S_1$ and
\qty{400}{\meter} for $S_2$. The authority would select $S_1$ although its target
ranks $S_2$ first. Each value is correct for the retained point but is a minimum,
not the declared mean. The reversal occurs because the evidence depends on the
plan; its direction depends on the metric and outcomes.

Let $\mathcal D=(x_i,g_i)_{i=1}^n$ be a fixed evaluation dataset, where
$x_i=(o_i,d_i)$ contains trip $i$'s origin and destination, $g_i$ is its proxy
group label, and $[n]=\{1,\ldots,n\}$. Let $a_i(S)$ be its combined endpoint
distance under
plan $S$, as defined in Equation~\ref{eq:access} below, and let
$y(S)=(a_i(S))_{i=1}^n$ be the resulting outcome vector. Before comparison, the
study fixes the target evidence
\begin{equation}
  \mathcal E^*=\bigl(J^*,\mathcal P^*,\mathcal C^*\bigr)
  \quad\text{and}\quad
  \theta(S)=F\bigl(y(S);\mathcal D,\mathcal E^*\bigr),
  \label{eq:target-estimand}
\end{equation}
where $J^*$ is the set of target observations, $\mathcal P^*$ is a reporting
partition, and $\mathcal C^*$ is an optional pool of candidate blocks. An
implemented audit $A$ may instead supply
\begin{equation}
  \mathcal E_A(S)=\bigl(J_A(S),\mathcal P_A(S),\mathcal C_A(S)\bigr),
  \label{eq:evidence-object}
\end{equation}
where the three components again specify observations, a partition, and a pool
of candidate blocks. A deterministic score function $F$ returns
\begin{equation}
  q_A(S)=F\bigl(y(S);\mathcal D,\mathcal E_A(S)\bigr).
  \label{eq:audit-score}
\end{equation}
Components unused by a score are empty. We say that the implementation conducts
an audit \emph{with common evidence for the stated target} when
$\mathcal E_A(S)=\mathcal E^*$ for every compared plan. The analyst may use
geometry and fixed proxy labels to construct $\mathcal E^*$, but not access
outcomes from individual plans.

\paragraph{Audit-consistency condition (by definition).}
For fixed $\mathcal D$, $F$, and target evidence $\mathcal E^*$, an implementation
with common evidence satisfies $q_A(S)=\theta(S)$ for every plan and therefore
\begin{equation}
  q_A(S)-q_A(S')=
  F\bigl(y(S);\mathcal D,\mathcal E^*\bigr)-
  F\bigl(y(S');\mathcal D,\mathcal E^*\bigr).
  \label{eq:common-difference}
\end{equation}
The equality follows directly by substituting
$\mathcal E_A(S)=\mathcal E_A(S')=\mathcal E^*$ into the two score definitions,
so only the outcome vector depends on the plan. Without common evidence or a
separately proved invariance condition, scores computed from plan-dependent
evidence may fail to reproduce the target ordering.
The example with two demand points uses the same
function $F$ to average the retained observations, but the retained set depends
on the plan. Its two reported values rank $S_1$ ahead of $S_2$ and do not recover
the target ordering, which ranks $S_2$ ahead of $S_1$.

Satisfying this condition ensures only that both scores represent the same
target and that their difference can be attributed to the outcome vectors. Causal or normative
fairness claims require additional assumptions. In the primary audit, we fix
$J^*=[n]$ and one partition. Section~\ref{sec:pmf} applies the same condition to
the secondary diagnostic.

\subsection{OD access and fixed proxy groups}

Let an evaluation split contain $n$ trips. Trip $i$ has origin
$o_i\in\mathbb{R}^2$, destination $d_i\in\mathbb{R}^2$, and proxy group label
$g_i\in G$ assigned from its origin area. The finite set $G$ contains all proxy
groups, and $U$ denotes the finite set of candidate sites. A station plan
$S\subseteq U$ has budget $|S|=k$. Its endpoint distance is
\begin{equation}
  a_i(S)=\min_{s\in S}\lVert o_i-s\rVert_2+
         \min_{s\in S}\lVert d_i-s\rVert_2.
  \label{eq:access}
\end{equation}
It sums origin access and destination egress but excludes travel between
stations; smaller values are better.
The evaluation artifact (the
saved outputs of each run) reports the overall mean, median, 90th percentile
(p90), 95th percentile (p95), and maximum, as well as these summaries
separately for origin and destination access. Let
$I_g=\{i:g_i=g\}$ contain the $n_g=|I_g|$ trips in group $g$. Write $Q_q(X)$ for
the $q$-quantile of a finite multiset $X$ and
$\bar a_g(S)=\frac{1}{n_g}\cdot\sum_{i\in I_g}a_i(S)$ for group $g$'s mean
endpoint distance. The two group metrics are
\begin{align}
  W_{90}(S)&=\max_{g\in G}\pNinety(\{a_i(S):i\in I_g\}),\\
  D_G(S)&=\max_g \bar a_g(S)-\min_g \bar a_g(S).
  \label{eq:group-metrics}
\end{align}
The worst-group p90, $W_{90}$, is the largest p90 among the groups, while $D_G$
is the difference between the highest and lowest group mean. They summarize tails and means across
the fixed proxy groups. Individual fairness would require a separate criterion
for each trip.

\subsection{Fixed grid constructed without plan outcomes}
\label{sec:fixed-grid}

We evaluate every plan on the same partition $\mathcal P^*$. We place a
square grid in the coordinate frame fitted to training geometry and fix its cell
size and origin before validation. We assign each validation or test trip to a
cell using only its
origin coordinates.

A cell is \emph{auditable} when it contains at least $m_{\min}$ trips. Let
$\mathcal P_{\mathrm{aud}}$ be the auditable cells and let $\bar a_p(S)$ be the mean
endpoint distance in cell $p$. The artifact reports both the full range of cell
means and the cell gap
\begin{equation}
  \cellgap(S)=Q_{0.9}(\{\bar a_p(S):p\in\mathcal P_{\mathrm{aud}}\})-
              Q_{0.1}(\{\bar a_p(S):p\in\mathcal P_{\mathrm{aud}}\}).
  \label{eq:unit-gap}
\end{equation}
The 90th--10th percentile difference reduces, but does not remove, the influence
of extreme cells.
The audit is valid only if at least two cells are auditable and, together,
contain a predeclared minimum share of all trips. Cell membership is
therefore invariant across plans. The artifact also reports differences between
groups within a cell when every group has enough observations; these sparse
within-cell diagnostics are descriptive only and do not enter the selection
rule. The results foreground the five selection
metrics in Equation~\ref{eq:selection-vector} and additionally report the trip
share in auditable cells for the selected plans.

For a fixed limit $L$ on endpoint distance, we further report
\begin{equation}
  \caprate(S)=\frac{1}{n}\cdot\sum_{i=1}^n \mathbf 1[a_i(S)\le L].
  \label{eq:absolute-access}
\end{equation}
Here $\mathbf 1[\cdot]$ equals one when its condition holds and zero otherwise.
$\caprate(S)$ is therefore the share of trips whose endpoint distance is
at most $L$.
All $n$ sampled trips are retained in the primary audit in
Equations~\ref{eq:access}--\ref{eq:absolute-access}; cell summaries use the
auditable cells defined above. Section~\ref{sec:setup} separately reports losses
during data sourcing and preprocessing.

\subsection{Selection rule fixed before validation}

We use the seven generators to build the plan portfolio from training data.
Because cell support is independent of the station plan, the audit-validity
condition in Section~\ref{sec:fixed-grid} applies to the entire split rather
than to individual plans. For a valid split, the predeclared lexicographic
selection rule minimizes the vector
\begin{equation}
  \bigl(W_{90}(S),\;\pNinety(\{a_i(S)\}_{i=1}^n),\;\cellgap(S),
  \;-\caprate(S),\;D_G(S)\bigr).
  \label{eq:selection-vector}
\end{equation}
The rule first minimizes the worst-group p90. It then breaks ties, in order, by
overall p90, cell gap, share within $L$, and group mean gap. To
describe trade-offs, we also identify the validation Pareto front: no other
compared plan is no worse in all five components and better in at least one.
With the same five components, a lexicographic minimizer is necessarily
nondominated, so restricting the implementation to that front leaves the choice
unchanged. Exact ties are broken by the lexicographic order of the method
identifier. This deterministic rule uses point estimates with no positive
indifference threshold; it does not establish statistical or practical
superiority. For prospective decisions, the contract should also specify how
practically indistinguishable scores are handled, for example through
predeclared indifference thresholds. We do not apply such thresholds
retrospectively. Other planning decisions may require a different metric
priority order.
These five audit metrics determine selection; PMF is secondary.

After writing the selected plan $S^*$ and the selection hash, the program loads
the test data and evaluates $S^*$ once. It does not compute test metrics for the other
six plans; selection therefore precedes all test metrics.

\section{Method}
\label{sec:method}

Algorithm~\ref{alg:protocol} applies the authority-supplied contract and metric
priorities from training through reporting.

\subsection{Seven methods for generating station plans}

The generators target spatial coverage, demand-weighted access, tail access by group, or
individual priority. The audit described above evaluates all resulting plans on
the same validation evidence.
All generators receive the same training trips, 300 candidate sites drawn from
endpoints observed in those trips, and budget $k=200$. This modeling choice
restricts siting to observed demand locations rather than independently
authorized sites; it makes the task closer to demand clustering than general
site selection. Its effect on the relative performance of Grid is not isolated.
The primary configuration has a constrained candidate universe: the large
$k/|U|=2/3$ ratio limits portfolio diversity, with pairs sharing 113 to 198
stations in Porto and 128 to 198 in Chicago. The post hoc check with
600 sites provides a complementary comparison with greater plan diversity
(Section~\ref{sec:sensitivity-results}).
We compute all seven plans before constructing the PMF pool:

\begin{enumerate}
  \item \textbf{Gonzalez} uses farthest-first traversal of $U$
  \cite{gonzalez1985kcenter}.
  \item \textbf{K-medoids} is a bounded $k$-medoids swap heuristic over
  candidate-site demand. Starting from $k$-means++ medoids, it applies the best
  improving swap in each of at most two exhaustive rounds.
  \item \textbf{Demand} applies a demand-weighted $p$-median objective and the
  same bounded swap search to all observed origin and destination demands.
  \item \textbf{Group-fair mean} minimizes the highest proxy-group mean endpoint
  distance, then the overall mean, so the group with the highest average distance
  has priority \cite{ghadiri2021socially}.
  \item \textbf{IFkCO} adapts the individually fair $k$-center with outliers
  objective to OD trips and applies a 10\%
  training outlier allowance. For each pair of trips, it compares two endpoint
  matchings: origin to origin and destination to destination, or the crossed
  matching. It chooses the matching with the smaller maximum endpoint distance,
  maps the representative endpoints to the nearest sites in $U$, and fills the
  remaining budget by farthest-first selection
  \cite{han2023ifkco}. The later audit nevertheless includes every validation and
  test trip.
  \item \textbf{Priority} applies a priority-center adaptation. It minimizes the
  worst ratio of a group's maximum endpoint distance to that group's p90 under
  the Gonzalez baseline plan,
  followed by the overall maximum and mean
  \cite{bajpai2021priority}.
  \item \textbf{Grid} is a control that places sites near the centroids of the
  grid cells containing the most trip endpoints and fills the budget by
  farthest-first. It uses
  endpoint counts and geometry alone.
\end{enumerate}

Group-fair mean, IFkCO, and Priority are adaptations for station placement;
Grid is a control.

\subsection{Secondary proportional mean fairlet diagnostic}
\label{sec:pmf}

We use this secondary diagnostic to study how a plan-dependent proposal pool
changes the basis of comparison; it does not select the station plan. A
\emph{fairlet} is a small mixed-group block that a later clustering step treats
as one unit. The diagnostic has three stages. First, we construct one common
pool of local blocks from trip origins and fixed proxy labels. Second, for each
station plan, we identify blocks whose group mean origin-access values differ
by at most $\varepsilon_m$. Third, the same set-packing model selects an
admissible collection of disjoint eligible blocks. The pool is shared across
plans; plan outcomes affect only block eligibility. We measure origin access
only, so equal means can reflect leveling down as well as improvement.

We fit the group template $t=(t_g)_{g\in G}$ from training group counts;
$t_g$ is the required number of group-$g$ trips in every block. Conceptually,
$\mathcal B^*$ contains all blocks that meet the geometric and compositional
requirements below. Because enumerating $\mathcal B^*$ is impractical, the
implementation constructs a deterministic proposal pool
$\mathcal C\subset\mathcal B^*$ before inspecting any station outcome.
Every block
$B\in\mathcal C$ has exact counts $|B\cap I_g|=t_g$ and a medoid radius at most
$R_{\max}$. Here the medoid is the block member with the smallest total
straight-line origin distance to the other members, and the radius is the
largest distance from that medoid. We reuse the same proposal pool for all seven
validation plans. After fixing the pool, each plan's outcomes determine which
blocks meet the threshold. For a plan $S$, let $\bar a^{\mathrm{orig}}_g(B,S)$ be the mean
distance from an origin to its nearest station among group $g$ trips in $B$, and
let $\mathcal C_S\subseteq\mathcal C$ be the eligible subset. A block
$B\in\mathcal C$ belongs to $\mathcal C_S$ when
\begin{equation}
  \max_{g,g'\in G}|\bar a^{\mathrm{orig}}_g(B,S)-
  \bar a^{\mathrm{orig}}_{g'}(B,S)|\le\varepsilon_m.
  \label{eq:pmf-gap}
\end{equation}
The fixed threshold $\varepsilon_m$ is measured in \si{\meter}; every plan uses
the same value.

For each $B\in\mathcal C_S$, binary variable $z_B$ indicates whether the model
selects that block. We solve the following set-packing model:
\begin{subequations}
\label{eq:pmf-packing-model}
\begin{align}
  \max\quad & M=\sum_{B\in\mathcal C_S}z_B,\\
  \text{s.t.}\quad & \sum_{B\in\mathcal C_S:\,i\in B}z_B\le1
      &&\forall i,\label{eq:packing}\\
  & M\ge M_{\min},\\
  & \left|M\cdot\left(\frac{b}{n}-\frac{t_g}{n_g}\right)\right|
      \le \delta_{\mathrm{out}} &&\forall g,\\
  & z_B\in\{0,1\} &&\forall B\in\mathcal C_S.\nonumber
\end{align}
\end{subequations}
Here $z_B=1$ means that block $B$ is selected, and the first constraint prevents
any trip from appearing in two selected blocks. We define
$b=\sum_g t_g$; $M$ is the number of selected blocks, and $M_{\min}$ is the
smallest block count that satisfies both the overall inclusion floor and every
group inclusion floor. With inclusion floor $\rho_{\mathrm{in}}$, a fixed
template gives
overall inclusion $\frac{M\cdot b}{n}$ and inclusion $\frac{M\cdot t_g}{n_g}$
for group $g$, so
$M_{\min}=\max\bigl(\lceil \rho_{\mathrm{in}} n/b\rceil,\,
\max_g\lceil \rho_{\mathrm{in}} n_g/t_g\rceil
\bigr)$.
The exclusion-rate constraint limits the \emph{exclusion-rate gap}, the largest absolute
difference between overall and group-specific exclusion rates, to
$\delta_{\mathrm{out}}$. Whenever $t_g/b\ne n_g/n$, the coefficient of $M$ in
group $g$'s exclusion-rate-gap constraint is nonzero, so that constraint can
also impose an \emph{upper} bound on $M$. Thus the inclusion floors, gap limit, and
packing constraints define an interval of
admissible block counts;
$M\ge M_{\min}$ alone is insufficient. A result is admissible only when this
interval contains the returned positive $M$.

SciPy/HiGHS solves the mixed-integer model in
Equation~\ref{eq:pmf-packing-model} with a deterministic 10{,}000-node
budget and zero requested relative gap. The incumbent is the best feasible
packing found, the upper bound limits the size of any eligible packing in
$\mathcal C_S$, and their relative difference is the mixed-integer-programming
(MIP) gap. We
also report node count, completion status, and maximum selected radius. We use
``optimal'' only relative to the eligible subset induced by the generated pool
$\mathcal C$.

\begin{algorithm}[t]
\caption{Freeze--validate--report protocol. PMF denotes proportional mean fairlet.}
\label{alg:protocol}
\begin{algorithmic}[1]
\REQUIRE splits $D_{\mathrm{tr}}$, $D_{\mathrm{val}}$, $D_{\mathrm{test}}$; station budget $k$; predeclared audit settings
\STATE Fix the frame and $U$ from $D_{\mathrm{tr}}$; record grid, groups, thresholds, and selection rule
\STATE Generate seven station plans from $D_{\mathrm{tr}}$ without PMF scores
\STATE Build one fairlet proposal pool on $D_{\mathrm{val}}$ without plan outcomes
\FOR{each plan $S$ generated from $D_{\mathrm{tr}}$}
  \STATE Compute the five audit metrics on the fixed grid for $D_{\mathrm{val}}$
  \STATE Solve the fairlet packing on the same pool for diagnosis only
\ENDFOR
\STATE Record the Pareto front; lexicographically select $S^*$; hash the decision
\STATE Load $D_{\mathrm{test}}$; report the audit metrics for $S^*$ exactly once
\STATE Build one test proposal pool without plan outcomes and report PMF for $S^*$
\RETURN selected plan, test audit, PMF result for the generated pool, manifests
\end{algorithmic}
\end{algorithm}

\section{Experimental Setup}
\label{sec:setup}

We instantiate the planning comparison retrospectively on two datasets with
fixed time splits, selection criteria, and deterministic search budgets.

\subsection{Temporal data and proxy groups}

Porto-SES uses public taxi trajectories joined by origin location to four Census
2021 socioeconomic strata \cite{moreiramatias2015porto,portugalCensus2021}.
Chicago-Hardship uses Transportation Network Provider trips joined by pickup
community area to four Hardship Index strata
\cite{chicagoTNP,chicagoHardship}. Both assignments describe a socioeconomic
score for the trip's origin area. The resulting proxy labels contain no traveler
attributes or verified residence data \cite{jacobs2021measurement}. The
Porto mobility records are from 2013 and the Census is from 2021; Chicago trips
are from 2022 and the Hardship Index source covers 2008--2012.

For Porto, we retain trajectories in each period with at least ten recorded
points and positive duration. After these filters, the three windows contain
9{,}815, 9{,}260,
and 9{,}081 trips for train, validation, and test. We sample 5{,}000 with seeds
701, 708, and 722. We spatially join origin points to Census sections. When a
point does not fall inside a section, we assign it to the nearest section; this
affects 2, 0, and 1 sampled points.
For Chicago, we require finite pickup and drop-off centroids, duration, trip
miles, and a Hardship score for the pickup community. Before sampling, 2{,}699,
4{,}762, and
3{,}395 rows are eligible; 0, 6, and 2 otherwise valid rows lack the score.
Seeds 701, 801, and 1001 select 2{,}500 per role. The metadata do not enumerate
every raw row rejected for malformed geometry or trip fields. Here, ``zero
exclusion'' means that no sampled trip is excluded from the primary audit.

Each prepared split stores coordinates in a local equirectangular frame centered
at its own mean origin, using Earth radius \qty{6371}{\kilo\meter}. Before
evaluation, we transform validation and test coordinates into the training
split's frame. All station distances and grid assignments therefore use one
training frame. We divide each dataset's area-level scores into quartiles. A higher
Porto label means higher constructed SES; a higher Chicago label means greater
hardship. Tables present the source-index quartiles in increasing order (very
low, low, middle, high). The quartiles are computed over areas, not sampled
trips; this explains the unequal group counts across the fixed splits.

We fixed all periods before comparing plans.
A gap separates validation from the later test window, which we used only after
developing the protocol. The stored run sequence declares the selection decision
before loading the test data.
The primary Chicago train, validation, and test windows are all on June 1,
2022: 07:00--08:00, 08:00--09:00, and 10:00--11:00. Clock time is therefore
confounded with split role in the primary design.

\begin{table*}[t]
\centering
\scriptsize
\tablelayout
\caption{Primary time splits and sampled counts for each proxy group. All Chicago intervals are on June 1, 2022, and every interval includes its start but not its end. VL/L/M/H denote very low, low, middle, and high source-index quartiles.}
\label{tab:splits}
\begin{tabular}{@{}lllrl@{}}
\toprule
Setting & Role & Period & $n$ & group counts VL/L/M/H \\
\midrule
Porto-SES & train & Jul 1--8, 2013 & $5{,}000$ & 1387/525/432/2656 \\
 & validation & Jul 8--15, 2013 & $5{,}000$ & 1423/521/494/2562 \\
 & test & Jul 22--29, 2013 & $5{,}000$ & 1508/501/454/2537 \\
\midrule
Chicago-Hardship & train & Jun 1, 07:00--08:00 & $2{,}500$ & 150/152/206/1992 \\
 & validation & Jun 1, 08:00--09:00 & $2{,}500$ & 159/193/196/1952 \\
 & test & Jun 1, 10:00--11:00 & $2{,}500$ & 127/126/191/2056 \\
\bottomrule
\end{tabular}
\end{table*}

\subsection{Selection and PMF parameters fixed before validation}

Porto-SES uses grid cells of width \qty{3}{\kilo\meter} and a limit
$L=\qty{2}{\kilo\meter}$ on endpoint distance. Chicago-Hardship uses
\qty{5}{\kilo\meter} cells and $L=\qty{5}{\kilo\meter}$. A cell needs 25 trips;
at least two cells and 90\% of
trips must be auditable. Grid origins are $(0,0)$ in each training coordinate
frame. The set of candidate sites built from training data has $|U|=300$, the
station budget is $k=200$, and bounded swap methods use at most two rounds.
The cells group nearby trip origins for the spatial comparison; $L$ determines
the reported share of trips within the stated endpoint-distance limit.

We also fix the secondary PMF parameters. Template entries below use display
order (very low, low, middle, high). Porto-SES uses block size 12, template
$(3,1,1,7)$, $\varepsilon_m=\qty{250}{\meter}$, and
$R_{\max}=\qty{3}{\kilo\meter}$. Chicago-Hardship uses block size 15, template
$(1,1,1,12)$, $\varepsilon_m=\qty{500}{\meter}$, and
$R_{\max}=\qty{5}{\kilo\meter}$. Both require
inclusion floors of $\rho_{\mathrm{in}}=0.25$ overall and for every group, and
$\delta_{\mathrm{out}}=0.30$. Pool construction uses seeds 101--105, three refill rounds,
and six attempts per seed. We fix search effort by these attempt counts instead
of wall-clock time.

\subsection{Ablation with plans and data held fixed}
\label{sec:ablation-setup}

We test whether using plan outcomes to construct the proposal pool changes the
PMF diagnostic while holding the seven validation plans and trips fixed. The
first design gives every plan the same stored pool, which uses only geometry and
group labels. The second design reproduces an earlier constructor from our own
pipeline. It builds a separate pool for each plan and uses origin access
when assigning points to block anchors and when checking the difference
between group means in each resulting
block.

Both designs use the same templates, limits on radius and differences between
group means, and requirements for inclusion and exclusion. They also use seeds
101--105, three refills, six attempts per seed, and at most 5{,}000 unique
proposed blocks. We give the same binary set-packing model a 10{,}000-node limit.
\Needspace{4\baselineskip}
The numbers of proposed blocks may differ because the two constructors
define different pools. For display only, we apply a fixed diagnostic order. It
first ranks admissible results ahead of unresolved and inadmissible results within
the pool. It then prefers higher minimum-group and overall inclusion, a lower
exclusion-rate gap, and more proposed blocks that meet the outcome threshold.
Exact ties are resolved by the lexicographic order of the method identifier.
Equation~\ref{eq:selection-vector}, not this diagnostic order, selects the station
plan. With one fixed template, these inclusion values and the exclusion-rate gap
are partly redundant functions of the selected block count. Tied plans receive the
same \emph{diagnostic rank}. Lower ranks among plans whose PMF results all fail
the fixed pool requirements have no substantive admissibility meaning.

Before generating results with a separate pool for each plan, we froze the
hypotheses, code and input hashes, phase log, and criteria for a relevant
difference: a change in PMF status or the top-ranked set, any strict pairwise
reversal, a rank shift of at least two, or Kendall $\tau_b<0.8$. We restrict the
ablation to validation; Section~\ref{sec:discussion} states the chronology limits.

\subsection{Post hoc sensitivity checks}
\label{sec:sensitivity-setup}

After inspecting the primary results, we specified two checks based on the five
audit metrics. The first expands the set of candidate sites from $|U|=300$ to
600. It holds $k=200$, all seven generators, primary time splits, grid,
thresholds, and selection rule fixed. The check tests how the high primary
$k/|U|$ ratio affects plan diversity and selection.

The Chicago check separates a change in calendar date from a change in hour. It
uses the same hour, 07:00--08:00, on three Wednesdays: June 1 for training, June 8 for
validation, and June 15 for the test used only for reporting. Seeds 701, 708, and
715 select 2{,}500 trips from 2{,}699, 3{,}578, and 3{,}727 eligible rows with a
Hardship Index. We retain $|U|=300$ and the validation selection rule. We
disable PMF because these checks address candidate-set size and temporal design.
Together,
the checks test whether the rule's selection depends on these two design
choices. Because we designed the checks post hoc, we do not treat them as
confirmatory or use them to estimate sampling uncertainty.

\paragraph{Post hoc bootstrap sensitivity.}
For the primary 300-site configurations, we additionally draw 10{,}000 bootstrap
samples per city from validation trips, resampling with replacement within each
proxy group while retaining its observed size. Every plan uses the same sampled
trip indices. In each draw we recompute all group p90 values using linear
interpolation and their maximum $W_{90}$; the worst group may change.
The 2.5th and 97.5th percentiles of the bootstrap distributions summarize
$W_{90}$ and paired differences. Plans, group labels and sizes, and the original
selection remain fixed; no test trips enter.
These exploratory ranges are neither simultaneous nor selection-adjusted.

\subsection{Reproducibility}

Each output directory stores source hashes; files listing candidate sites and
stations in comma-separated values (CSV) format; the fixed grid; validation and test records;
the selection hash; proposal pools; and solver results. A phase log with seven
recorded events shows that the program records the validation selection before
loading the test data.
A separately implemented verifier reconstructs candidate sites, audit
metrics, selection, PMF pools, block templates, radii, differences between group
means, inclusion and exclusion requirements, bounds, and node budgets. It
passes 38 checks on the main evaluation artifact. A separate ablation verifier
checks the recorded decision order, absence of test inputs, source and plan
hashes, validity of candidate sites, and saved packings. It also checks saved access
outcomes, ranks, ties, reversals, and flags for the stated criteria. All 1{,}570 ablation
checks pass. The artifact for the expanded set of candidate sites passes 20 checks, and
the Chicago artifact for the same hour across dates passes 11.
The bootstrap supplement stores its configuration, seeds, hashes, all draws,
and all 21 pairwise differences per city; 11 synthetic tests and 335 separate
checks of output arithmetic and input integrity pass.
These separately implemented verifiers reproduce the saved computations. The
results have not been independently replicated by an external group.

\section{Results}
\label{sec:results}

\begin{table*}[t]
\centering
\scriptsize
\tablelayout
\caption{Validation metrics in selection-rule order. Arrows show the preferred direction; bold marks the selected plan and ``yes'' a nondominated plan. Within $L$ is the share of trips within the endpoint-distance limit.}
\label{tab:validation}
\begin{tabular}{@{}llrrrrrc@{}}
\toprule
Setting & Plan generated on training & worst-group p90 (\si{\meter})$\downarrow$ & overall p90 (\si{\meter})$\downarrow$ & cell gap (\si{\meter})$\downarrow$ & within $L$ \%$\uparrow$ & group mean gap (\si{\meter})$\downarrow$ & Pareto? \\
\midrule
Porto-SES & Gonzalez & 1{,}404 & 1{,}259 & 222 & 99.7 & 203 & no \\
 & K-medoids & 1{,}400 & 1{,}269 & 189 & 99.5 & 250 & no \\
 & Demand & 1{,}239 & 1{,}075 & 101 & 99.6 & 222 & yes \\
 & Group-fair mean & 1{,}238 & 1{,}127 & 108 & 99.6 & 141 & yes \\
 & \textbf{IFkCO} & 885 & 805 & 319 & 97.7 & 180 & yes \\
 & Priority & 1{,}224 & 1{,}107 & 201 & 99.7 & 175 & yes \\
 & Grid & 1{,}186 & 1{,}145 & 180 & 99.6 & 132 & yes \\
\midrule
Chicago-Hardship & Gonzalez & 1{,}822 & 1{,}508 & 581 & 100.0 & 388 & yes \\
 & K-medoids & 1{,}917 & 1{,}732 & 499 & 100.0 & 160 & yes \\
 & Demand & 1{,}857 & 1{,}422 & 637 & 100.0 & 458 & yes \\
 & Group-fair mean & 1{,}876 & 1{,}508 & 599 & 100.0 & 403 & yes \\
 & IFkCO & 11{,}547 & 1{,}272 & 1{,}894 & 96.4 & 3{,}618 & yes \\
 & Priority & 1{,}757 & 1{,}347 & 647 & 100.0 & 511 & yes \\
 & \textbf{Grid} & 1{,}735 & 1{,}442 & 567 & 100.0 & 423 & yes
\\
\bottomrule
\end{tabular}
\end{table*}

\subsection{Validation comparison and selection}

The validation comparison reports the five audit metrics used in
Equation~\ref{eq:selection-vector}. Both settings meet the requirements on the
number and trip share of auditable cells, and the audit includes every trip for
every plan.
Five of seven Porto-SES plans are nondominated. IFkCO uniquely minimizes the
first-priority metric, with a worst-group p90 of \qty{885}{\meter}, so no
lower-priority metric is consulted. Its 97.7\% share within the access limit and
its cell gap are worse than those of several alternatives. The Pareto label only
describes trade-offs.

All seven Chicago-Hardship plans are nondominated. IFkCO has the lowest
overall p90 but a
worst-group p90 above \qty{11}{\kilo\meter}, whereas Grid uniquely minimizes the
first-priority metric, by \qty{22}{\meter} over the next-best plan. The
predeclared rule therefore chooses Grid, whose worst-group p90 is
\qty{1735}{\meter}. Its \qty{22}{\meter} advantage does not establish
statistically reliable or practically meaningful superiority over Priority.
Overall p90 and
worst-group p90 favor different plans.
\Needspace{4\baselineskip}
With
validation groups ranging from 159 to 1{,}952 trips, about 16
observations lie in the smallest group's upper decile. The reported $W_{90}$
is a tail point estimate and can change with the sample.
Table~\ref{tab:bootstrap} reports the corresponding post hoc sensitivity.

\begin{figure*}[t]
\centering
\begin{minipage}[t]{0.49\textwidth}
  \centering
  \includegraphics[width=0.96\linewidth]{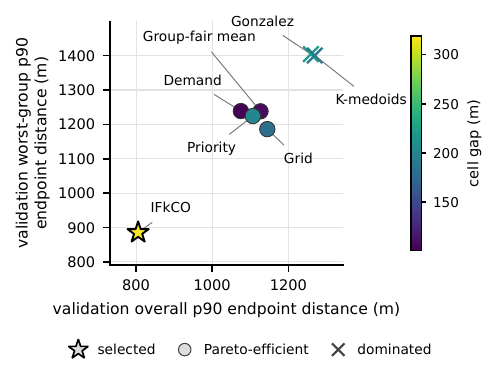}
  \centerline{(a) Porto-SES}
\end{minipage}\hfill
\begin{minipage}[t]{0.49\textwidth}
  \centering
  \includegraphics[width=0.96\linewidth]{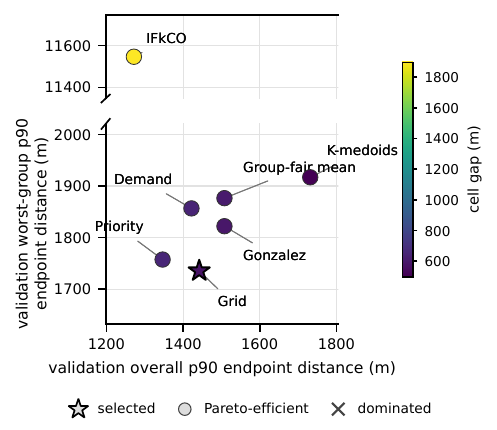}
  \centerline{(b) Chicago-Hardship}
\end{minipage}
\caption{Validation overall versus worst-group p90 endpoint distance. Stars mark
selected plans, circles other Pareto plans, and crosses dominated plans (legend
below each panel); color encodes the cell gap (colorbar). The Chicago panel's
vertical axis is broken so IFkCO's outlying worst-group value and the other six
plans both stay legible. Table~\ref{tab:validation} gives the reported values and all
five selection metrics.}
\Description{Two scatter plots compare seven station plans on validation overall and worst-group p90 endpoint distance, each with a colorbar for cell gap and a legend for selected, Pareto, and dominated markers. Porto-SES selects IFkCO; Chicago-Hardship selects Grid. The Chicago panel's y-axis is broken so IFkCO's outlying worst-group value and the other six plans are both legible.}
\label{fig:pareto}
\end{figure*}

\subsection{Test results without reselection}

We compare validation and test values for each selected plan; test data never
affect selection. Porto-SES overall p90 remains \qty{805}{\meter}, while
worst-group p90 changes from \qty{885}{\meter} to \qty{872}{\meter} and the
group mean gap narrows from \qty{180}{\meter} to \qty{174}{\meter}. For
Chicago-Hardship, overall p90 changes from \qty{1442}{\meter} to
\qty{1404}{\meter}, worst-group p90 changes from \qty{1735}{\meter} to
\qty{1656}{\meter}, and the group mean gap narrows from \qty{423}{\meter} to
\qty{314}{\meter}. These are descriptive point estimates. In Chicago,
the change in clock time also permits a change in rider composition, so these
differences do not isolate temporal generalization from changes in demand.

\Needspace{6\baselineskip}
The Porto-SES and Chicago-Hardship test cell gaps are \qty{161}{\meter} and
\qty{521}{\meter}, down from \qty{319}{\meter} and \qty{567}{\meter} on
validation. Auditable cells contain 98.4\%/98.7\% (validation/test) of trips on
Porto-SES and 95.7\%/95.9\% on Chicago-Hardship, both above the fixed 90\%
floor. The respective shares within the access limit are 97.7\% at
\qty{2}{\kilo\meter} and 100\% at \qty{5}{\kilo\meter}, unchanged between
splits. Cell support affects only the cell gap; every trip enters the other
metrics. Because Porto has only eight auditable cells in both splits, its cell
gap remains sensitive to individual cells; Chicago has 16 on validation and 15
on test.
Together, the validation comparison and frozen test report answer the research
question for the primary audit: the planning authority can trace the choice to
one validation evidence set shared by all seven plans, and the selected plan is
fixed before the test data are loaded.

\begin{table*}[t]
\centering
\scriptsize
\tablelayout
\caption{PMF validation ranks under a common pool and one pool per plan (1 =
best; tied plans share a rank; $^\ast$ marks the display choice after tie-breaking). Switching
from the common pool to one pool per plan reverses Grid and K-medoids in Porto
and changes the Chicago ties. All Chicago results are inadmissible; their ranks
describe only the diagnostic ordering, not admissible alternatives.
The selection rule in Equation~\ref{eq:selection-vector}, not the PMF
diagnostic, selects the station plan.}
\label{tab:audit-design-ranks}
\begin{tabular}{@{}llccccccc@{}}
\toprule
Setting & Pool & IFkCO & Priority & Group-fair mean & Demand & Grid & K-medoids & Gonzalez \\
\midrule
Porto-SES & common pool & \textbf{1$^\ast$} & 2 & 3 & 4 & 5 & 6 & 7 \\
 & pool per plan & \textbf{1$^\ast$} & 2 & 3 & 4 & 6 & 5 & 7 \\
\midrule
Chicago-Hardship & common pool & 1 & 6 & 4 & 4 & \textbf{1$^\ast$} & 3 & 6 \\
 & pool per plan & 2 & 4 & 4 & 4 & \textbf{1$^\ast$} & 3 & 4
\\
\bottomrule
\end{tabular}
\end{table*}

\subsection{Post hoc sensitivity checks}
\label{sec:sensitivity-results}
Table~\ref{tab:candidate-sensitivity} contrasts the constrained primary case
with the expanded candidate set: minimum overlap falls and selection changes
in both cities. The largest
pairwise overlap remains 199 stations in each city: Demand and Group-fair mean
in Porto, and Demand and K-medoids in Chicago, for $|U|=600$.
Thus the more diverse configuration does not
support a generally preferred generator inferred from the constrained 300-site case. The
expanded case was specified after inspecting the primary results and remains
exploratory.

\begin{table}[htbp]
\centering
\footnotesize
\tablelayout
\caption{Candidate-set sensitivity at fixed $k=200$. Minimum overlap is the
smallest number of shared stations across plan pairs. The 300-site cases are
primary; the 600-site cases are post hoc. Selection uses validation only.}
\label{tab:candidate-sensitivity}
\begin{tabular}{@{}lrrl@{}}
\toprule
Setting & $|U|$ & Min. overlap & Selected generator \\
\midrule
Porto-SES & 300 & 113 & IFkCO \\
 & 600 & 39 & Grid \\
\midrule
Chicago-Hardship & 300 & 128 & Grid \\
 & 600 & 42 & Priority \\
\bottomrule
\end{tabular}
\end{table}

In the Chicago check at the same hour on different dates, the rule selects Priority on the
June 8 validation data. For this selected plan, the June
15 test gives
overall p90 \qty{1195}{\meter}, worst-group p90 \qty{1640}{\meter}, cell gap
\qty{463}{\meter}, 100\% within the access limit, and 94.0\% auditable-cell share over
12 auditable cells. Holding the hour fixed removes the primary design's
time-of-day difference from this check; pairs of plans still share 128--198
stations.
These changes in selection show why the authorized candidate sites and the
evaluation period are planning inputs rather than implementation details.

\paragraph{Sensitivity to resampled validation trips.}
Table~\ref{tab:bootstrap} compares each originally selected plan with the plan
having the next-lowest validation $W_{90}$. Chicago's range spans both orderings; Porto's remains
below zero for IFkCO minus Grid. This is a pairwise sensitivity result, not a
portfolio-wide superiority claim, and does not alter either selection.

\begin{table}[htbp]
\centering
\footnotesize
\tablelayout
\caption{Post hoc paired differences in validation $W_{90}$ (meters).
Negative values favor the first plan. Ranges are individual 2.5th--97.5th
bootstrap percentiles, without multiple-comparison adjustment.}
\label{tab:bootstrap}
\begin{tabular}{@{}llrr@{}}
\toprule
Setting & Difference & Observed & 95\% range \\
\midrule
Chicago & Grid -- Priority & $-22$ & $[-111, 229]$ \\
Porto & IFkCO -- Grid & $-301$ & $[-359, -30]$
\\
\bottomrule
\end{tabular}
\end{table}

\subsection{Secondary PMF results, bounds, and pool ablation}

For the selected Porto-SES plan on the test data and the block template
$(3,1,1,7)$ in display order (very low, low, middle, high), 777 of $1{,}387$
blocks constructed without plan outcomes pass the \qty{250}{\meter} threshold.
The 10{,}000-node
search returns 174 disjoint blocks with upper bound 184 and a 5.7\% gap,
reaching the deterministic node limit before the gap closes. The selected blocks
cover 41.8\% of trips; the lowest group inclusion rate is 34.6\%, the
exclusion-rate
gap is 0.071, and the radius is at most \qty{2.99}{\kilo\meter}. The result
meets all fixed requirements; the reported gap leaves optimality unresolved.

The Chicago-Hardship test pool, template $(1,1,1,12)$, contains 34 blocks; one
passes the \qty{500}{\meter} threshold. The inclusion floors require 42 blocks
overall and 48 for the most restrictive group, already exceeding the single
eligible block. After we impose these requirements, HiGHS classifies the model
as infeasible and returns no incumbent, so no upper bound, gap, or radius is
reported. This result establishes inadmissibility only inside the sparse
proposal pool; we do not test PMF feasibility over the complete set of possible
blocks. The diagnostic therefore supplies no admissible PMF solution for this
Chicago case, rather than evidence of fair access or general PMF infeasibility.
The primary plan audit is unchanged.

\paragraph{Common versus plan-specific pools.}
\label{sec:ablation-results}

\Needspace{8\baselineskip}
Table~\ref{tab:audit-design-ranks} reports the PMF diagnostic ranks under the
order fixed for this
ablation. In Porto-SES, the common pool contains 1{,}168 proposed
blocks; the seven pools built separately for each plan contain 291--783.
IFkCO has the unique top diagnostic rank and is the only plan with an admissible
PMF result under both designs.
The rankings have Kendall $\tau_b=0.905$ and Spearman $\rho=0.964$. Grid and
K-medoids reverse fifth and sixth place; the maximum shift is one.

\Needspace{5\baselineskip}
Chicago-Hardship has 16 proposed blocks in the common pool, while its separate
pools contain zero to four; every PMF result is inadmissible under the fixed pool
requirements in both designs.
Five ties change without a strict reversal; the largest rank shift is two,
$\tau_b=0.852$, and $\rho=0.911$. These ranks do not distinguish admissible
alternatives in Chicago.

The ablation shows limited changes in diagnostic ranks across pool designs:
Porto retains the same top-ranked admissible PMF result, and Chicago has no
admissible result under either design. It nevertheless illustrates why PMF
comparisons require a common pool.
Although the underlying trip set remains fixed, plan-specific constructors
generate different proposal pools, so the packing models optimize over
different feasible sets and the resulting PMF values lack a common basis of
comparison. For this ablation, the solver certifies all saved pool maxima as
optimal within their generated pools, including empty packings. These maxima
omit the inclusion and exclusion-rate constraints; admissibility is checked
separately.

\FloatBarrier
\section{Discussion and Limitations}
\label{sec:discussion}

\subsection{Validity controls}

The contract makes comparisons traceable to shared observations, reporting
cells, and proposal pools. Separating plan generation, validation selection,
and test reporting prevents test outcomes from influencing the recorded choice.
PMF checks inclusion, locality, and block composition and reports solver
bounds; fairness of the station plan remains a policy judgment.
The pool constructor changes PMF ranks, while candidate
sites and temporal design change plan selection. Each change can be traced to
an input declared in the contract.

\subsection{Prospective use in urban decision support}

For prospective use, a planning authority publishes a versioned contract before
receiving proposals.
\Needspace{6\baselineskip}
It specifies the target population and data cutoff, proxy
definitions, authorized sites, reporting grid and support rules, access metric
and limits, validation period, selection rule, and any proposal pools. Each
proposer supplies only a station set and plan hash. An analyst runs the same
program for every proposal and publishes validation vectors, Pareto labels,
audit validity, PMF admissibility, and hashes; an auditor checks the record
without redefining the evidence. The authority owns the normative inputs,
decides, may monitor later observations, and starts a new versioned review if an
input changes.

An authority can favor a plan by choosing candidate sites, evaluation periods,
or metric priorities after inspecting outcomes, even when all plans share the
same evidence. A prospective process should publish an externally
timestamped contract before comparison and document the reasons for subsequent
changes. Hashes alone do not establish this prior commitment.

This is a prospective workflow design; we have not obtained municipal or
stakeholder validation for the present data, limits, or selection priorities.
Deployment would require
authorized sites, access measured on the relevant transport network, policy
limits reviewed by stakeholders, repeated time windows, and a public archival
record.

\subsection{Comparative evaluation for urban decision intelligence}

Urban decision intelligence must compare alternatives from different method
families. The contract supplies a common evaluation layer for classical,
learned, and hybrid generators. If a system proposes both an intervention and
the units, blocks, or trips used to score it, Section~\ref{sec:common-evidence}
states what must remain common. Our fixed generators isolate this comparison
question; we make no empirical claim about learned generators.

\subsection{Remaining limitations}

The protocol records groups, thresholds, and metric priorities but does not
choose them for a planning authority. We do not compare alternative proxy
definitions, thresholds, or metric-priority orders, nor do we estimate decision
regret, training variability, or uncertainty for learned
generators. The area-level Porto-SES and
Chicago-Hardship proxies use source years that differ from the trip years, and
their strata contain unequal numbers of sampled trips. They describe origin
areas, not travelers' SES
\cite{jacobs2021measurement}. A fixed grid prevents plan-dependent aggregation,
but its scale and zoning remain subject to MAUP \cite{javanmard2023maup}.

The sampled trips represent observed customers, not demand for stations in the
full population. Candidate sites are trip endpoints rather than authorized parcels or
stops. Straight-line endpoint distance measures geometric proximity, not the
network, opportunities, schedules, fares, capacity, safety, or other dimensions
of accessibility \cite{geurs2004accessibility}. The bounded generators lack
approximation guarantees in several cases; we did not test the original
algorithm families.

The bootstrap assumes independent, exchangeable trips within proxy groups and
conditions on the observed group counts, plans, and windows. It does not capture
spatial or temporal dependence, changing group shares, or training uncertainty;
no finite-population correction is applied. We provide no rolling-origin
evaluation \cite{roberts2017crossvalidation}.
Chicago's \qty{22}{\meter} validation margin rests on a small group's tail, and
its primary windows confound split role with clock time. The same-hour
check holds clock time fixed in one post hoc sequence; it does not resolve the
confound in the primary comparison. Expanding the candidate
set changes both selections. Stored inputs and code reproduce the estimates, not
their sampling reliability or prospective generalization.

The common PMF pool contains only a sampled subset of all possible blocks, so
even a zero MIP gap would certify optimality only within that pool. Porto-SES
has a nonzero gap,
and Chicago-Hardship fails the pool requirements. A later clustering hierarchy
would need a predeclared rule that preserves the reported radius or diameter
limit at the selected cut.

Hashes verify file identity, and phase logs record the declared execution
order, but the chronology has no
external timestamp or public preregistration. We saw the
original test before the ablation, sensitivity checks, and bootstrap analysis;
the constructor comparison and bootstrap use validation only, and the added
analyses are labeled post hoc.
The package has no archival DOI at the time of writing.

\section{Conclusion}
\label{sec:conclusion}

A frozen evaluation contract lets a planning authority attribute score
differences to plan outcomes within one declared comparison. In both
retrospective cases, changing candidate sites or temporal design changes the
selected plan. Rather than remove this dependence, the protocol exposes it
before the authority acts. Its value lies in making populations, sites,
reporting cells, thresholds, and metric priorities explicit in the decision record.
PMF conclusions remain limited to sampled pools. A planning authority should
publish the contract and start a new evaluation whenever an input changes.

\begin{acks}
During the preparation of this work, the authors used OpenAI ChatGPT for
language editing and OpenAI Codex for Python implementation, test drafting,
artifact verification, and plotting. The authors reviewed and edited the
generated material and take full responsibility for the content and reported
results of the publication.
\end{acks}

\phantomsection\label{page:main-end}
\balance
\phantomsection\label{page:references-start}
\bibliographystyle{ACM-Reference-Format}
\bibliography{paper_refs}


\begin{thebibliography}{23}


\ifx \showCODEN    \undefined \def \showCODEN     #1{\unskip}     \fi
\ifx \showISBNx    \undefined \def \showISBNx     #1{\unskip}     \fi
\ifx \showISBNxiii \undefined \def \showISBNxiii  #1{\unskip}     \fi
\ifx \showISSN     \undefined \def \showISSN      #1{\unskip}     \fi
\ifx \showLCCN     \undefined \def \showLCCN      #1{\unskip}     \fi
\ifx \shownote     \undefined \def \shownote      #1{#1}          \fi
\ifx \showarticletitle \undefined \def \showarticletitle #1{#1}   \fi
\ifx \showURL      \undefined \def \showURL       {\relax}        \fi
\providecommand\bibfield[2]{#2}
\providecommand\bibinfo[2]{#2}
\providecommand\natexlab[1]{#1}
\providecommand\showeprint[2][]{arXiv:#2}
\makeatletter
\@ifundefined{NAT@parse@date}{}{\let\NAT@parse@date@orig\NAT@parse@date}
\@ifundefined{NAT@parse@date}{}{\def\NAT@parse@date#1#2#3#4#5#6@@{\NAT@parse@date@orig#1#2#3#4#5#6@@\def\NAT@tempyear{0000}\def\NAT@tempexlab{{?}}\ifx\NAT@year\NAT@tempyear\ifx\NAT@exlab\NAT@tempexlab\def\NAT@date{[n.\,d.]}\else\edef\NAT@date{[n.\,d.]\NAT@exlab}\fi\fi}}
\makeatother

\bibitem[Almanza et~al\mbox{.}(2022)]%
        {almanza2022fairoutliers}
\bibfield{author}{\bibinfo{person}{Matteo Almanza}, \bibinfo{person}{Alessandro
  Epasto}, \bibinfo{person}{Alessandro Panconesi}, {and}
  \bibinfo{person}{Giuseppe Re}.} \bibinfo{year}{2022}\natexlab{}.
\newblock \showarticletitle{{$k$}-Clustering with Fair Outliers}. In
  \bibinfo{booktitle}{\emph{Proceedings of the Fifteenth ACM International
  Conference on Web Search and Data Mining}}. \bibinfo{publisher}{Association
  for Computing Machinery}, \bibinfo{address}{New York, NY, USA},
  \bibinfo{pages}{5--15}.
\newblock
\href{https://doi.org/10.1145/3488560.3498485}{doi:\nolinkurl{10.1145/3488560.3498485}}


\bibitem[Bajpai et~al\mbox{.}(2021)]%
        {bajpai2021priority}
\bibfield{author}{\bibinfo{person}{Tanvi Bajpai}, \bibinfo{person}{Deeparnab
  Chakrabarty}, \bibinfo{person}{Chandra Chekuri}, {and}
  \bibinfo{person}{Maryam Negahbani}.} \bibinfo{year}{2021}\natexlab{}.
\newblock \showarticletitle{Revisiting Priority {$k$}-Center: Fairness and
  Outliers}. In \bibinfo{booktitle}{\emph{48th International Colloquium on
  Automata, Languages, and Programming (ICALP 2021)}}
  \emph{(\bibinfo{series}{Leibniz International Proceedings in Informatics},
  Vol.~\bibinfo{volume}{198})}. \bibinfo{publisher}{Schloss
  Dagstuhl--Leibniz-Zentrum f{\"u}r Informatik}, \bibinfo{address}{Dagstuhl,
  Germany}, \bibinfo{pages}{21:1--21:20}.
\newblock
\href{https://doi.org/10.4230/LIPIcs.ICALP.2021.21}{doi:\nolinkurl{10.4230/LIPIcs.ICALP.2021.21}}


\bibitem[Chierichetti et~al\mbox{.}(2017)]%
        {chierichetti2017fairlets}
\bibfield{author}{\bibinfo{person}{Flavio Chierichetti}, \bibinfo{person}{Ravi
  Kumar}, \bibinfo{person}{Silvio Lattanzi}, {and} \bibinfo{person}{Sergei
  Vassilvitskii}.} \bibinfo{year}{2017}\natexlab{}.
\newblock \showarticletitle{Fair Clustering Through Fairlets}. In
  \bibinfo{booktitle}{\emph{Advances in Neural Information Processing Systems
  30}}. \bibinfo{publisher}{Curran Associates, Inc.}, \bibinfo{address}{Red
  Hook, NY, USA}, \bibinfo{pages}{5029--5037}.
\newblock
\urldef\tempurl%
\url{https://proceedings.neurips.cc/paper/2017/hash/978fce5bcc4eccc88ad48ce3914124a2-Abstract.html}
\showURL{%
\tempurl}


\bibitem[{City of Chicago}(2014)]%
        {chicagoHardship}
\bibfield{author}{\bibinfo{person}{{City of Chicago}}.}
  \bibinfo{year}{2014}\natexlab{}.
\newblock \bibinfo{title}{Census Data -- Selected Socioeconomic Indicators in
  Chicago, 2008--2012}.
\newblock \bibinfo{howpublished}{Chicago Data Portal}.
\newblock
\urldef\tempurl%
\url{https://data.cityofchicago.org/Health-Human-Services/Census-Data-Selected-socioeconomic-indicators-in-C/jcxq-k9xf}
\showURL{%
\tempurl}
\newblock
\shownote{Includes the community-area Hardship Index; accessed 21 August 2026}.


\bibitem[{City of Chicago}(2023)]%
        {chicagoTNP}
\bibfield{author}{\bibinfo{person}{{City of Chicago}}.}
  \bibinfo{year}{2023}\natexlab{}.
\newblock \bibinfo{title}{Transportation Network Providers -- Trips
  (2018--2022)}.
\newblock \bibinfo{howpublished}{Chicago Data Portal}.
\newblock
\urldef\tempurl%
\url{https://data.cityofchicago.org/Transportation/Transportation-Network-Providers-Trips-2018-2022-/m6dm-c72p}
\showURL{%
\tempurl}
\newblock
\shownote{Accessed 21 August 2026}.


\bibitem[Dwork et~al\mbox{.}(2015)]%
        {dwork2015reusable}
\bibfield{author}{\bibinfo{person}{Cynthia Dwork}, \bibinfo{person}{Vitaly
  Feldman}, \bibinfo{person}{Moritz Hardt}, \bibinfo{person}{Toniann Pitassi},
  \bibinfo{person}{Omer Reingold}, {and} \bibinfo{person}{Aaron Roth}.}
  \bibinfo{year}{2015}\natexlab{}.
\newblock \showarticletitle{The Reusable Holdout: Preserving Validity in
  Adaptive Data Analysis}.
\newblock \bibinfo{journal}{\emph{Science}} \bibinfo{volume}{349},
  \bibinfo{number}{6248} (\bibinfo{year}{2015}), \bibinfo{pages}{636--638}.
\newblock
\href{https://doi.org/10.1126/science.aaa9375}{doi:\nolinkurl{10.1126/science.aaa9375}}


\bibitem[Geurs and van Wee(2004)]%
        {geurs2004accessibility}
\bibfield{author}{\bibinfo{person}{Karst~T. Geurs} {and} \bibinfo{person}{Bert
  van Wee}.} \bibinfo{year}{2004}\natexlab{}.
\newblock \showarticletitle{Accessibility Evaluation of Land-Use and Transport
  Strategies: Review and Research Directions}.
\newblock \bibinfo{journal}{\emph{Journal of Transport Geography}}
  \bibinfo{volume}{12}, \bibinfo{number}{2} (\bibinfo{year}{2004}),
  \bibinfo{pages}{127--140}.
\newblock
\href{https://doi.org/10.1016/j.jtrangeo.2003.10.005}{doi:\nolinkurl{10.1016/j.jtrangeo.2003.10.005}}


\bibitem[Ghadiri et~al\mbox{.}(2021)]%
        {ghadiri2021socially}
\bibfield{author}{\bibinfo{person}{Mehrdad Ghadiri}, \bibinfo{person}{Samira
  Samadi}, {and} \bibinfo{person}{Santosh Vempala}.}
  \bibinfo{year}{2021}\natexlab{}.
\newblock \showarticletitle{Socially Fair {$k$}-Means Clustering}. In
  \bibinfo{booktitle}{\emph{Proceedings of the 2021 ACM Conference on Fairness,
  Accountability, and Transparency}}. \bibinfo{publisher}{Association for
  Computing Machinery}, \bibinfo{address}{New York, NY, USA},
  \bibinfo{pages}{438--448}.
\newblock
\href{https://doi.org/10.1145/3442188.3445906}{doi:\nolinkurl{10.1145/3442188.3445906}}


\bibitem[Gonzalez(1985)]%
        {gonzalez1985kcenter}
\bibfield{author}{\bibinfo{person}{Teofilo~F. Gonzalez}.}
  \bibinfo{year}{1985}\natexlab{}.
\newblock \showarticletitle{Clustering to Minimize the Maximum Intercluster
  Distance}.
\newblock \bibinfo{journal}{\emph{Theoretical Computer Science}}
  \bibinfo{volume}{38} (\bibinfo{year}{1985}), \bibinfo{pages}{293--306}.
\newblock
\href{https://doi.org/10.1016/0304-3975(85)90224-5}{doi:\nolinkurl{10.1016/0304-3975(85)90224-5}}


\bibitem[Han et~al\mbox{.}(2023)]%
        {han2023ifkco}
\bibfield{author}{\bibinfo{person}{Lu Han}, \bibinfo{person}{Dachuan Xu},
  \bibinfo{person}{Yicheng Xu}, {and} \bibinfo{person}{Ping Yang}.}
  \bibinfo{year}{2023}\natexlab{}.
\newblock \showarticletitle{Approximation Algorithms for the Individually Fair
  {$k$}-Center with Outliers}.
\newblock \bibinfo{journal}{\emph{Journal of Global Optimization}}
  \bibinfo{volume}{87}, \bibinfo{number}{2--4} (\bibinfo{year}{2023}),
  \bibinfo{pages}{603--618}.
\newblock
\href{https://doi.org/10.1007/s10898-022-01195-3}{doi:\nolinkurl{10.1007/s10898-022-01195-3}}


\bibitem[{Instituto Nacional de Estat{\'i}stica}(2021)]%
        {portugalCensus2021}
\bibfield{author}{\bibinfo{person}{{Instituto Nacional de Estat{\'i}stica}}.}
  \bibinfo{year}{2021}\natexlab{}.
\newblock \bibinfo{title}{Censos 2021: {BGRI}, {GRID} e Lugares}.
\newblock \bibinfo{howpublished}{Geographic information download portal}.
\newblock
\urldef\tempurl%
\url{https://mapas.ine.pt/download/index2021.phtml}
\showURL{%
\tempurl}
\newblock
\shownote{Accessed 21 August 2026}.


\bibitem[Jacobs and Wallach(2021)]%
        {jacobs2021measurement}
\bibfield{author}{\bibinfo{person}{Abigail~Z. Jacobs} {and}
  \bibinfo{person}{Hanna Wallach}.} \bibinfo{year}{2021}\natexlab{}.
\newblock \showarticletitle{Measurement and Fairness}. In
  \bibinfo{booktitle}{\emph{Proceedings of the 2021 ACM Conference on Fairness,
  Accountability, and Transparency}}. \bibinfo{publisher}{Association for
  Computing Machinery}, \bibinfo{address}{New York, NY, USA},
  \bibinfo{pages}{375--385}.
\newblock
\href{https://doi.org/10.1145/3442188.3445901}{doi:\nolinkurl{10.1145/3442188.3445901}}


\bibitem[Javanmard et~al\mbox{.}(2023)]%
        {javanmard2023maup}
\bibfield{author}{\bibinfo{person}{Reyhane Javanmard},
  \bibinfo{person}{Jinhyung Lee}, \bibinfo{person}{Junghwan Kim},
  \bibinfo{person}{Luyu Liu}, {and} \bibinfo{person}{Ehab Diab}.}
  \bibinfo{year}{2023}\natexlab{}.
\newblock \showarticletitle{The Impacts of the Modifiable Areal Unit Problem
  ({MAUP}) on Social Equity Analysis of Public Transit Reliability}.
\newblock \bibinfo{journal}{\emph{Journal of Transport Geography}}
  \bibinfo{volume}{106} (\bibinfo{year}{2023}), \bibinfo{pages}{103500}.
\newblock
\href{https://doi.org/10.1016/j.jtrangeo.2022.103500}{doi:\nolinkurl{10.1016/j.jtrangeo.2022.103500}}


\bibitem[Jung et~al\mbox{.}(2020)]%
        {jung2020service}
\bibfield{author}{\bibinfo{person}{Christopher Jung}, \bibinfo{person}{Sampath
  Kannan}, {and} \bibinfo{person}{Neil Lutz}.} \bibinfo{year}{2020}\natexlab{}.
\newblock \showarticletitle{Service in Your Neighborhood: Fairness in Center
  Location}. In \bibinfo{booktitle}{\emph{1st Symposium on Foundations of
  Responsible Computing (FORC 2020)}} \emph{(\bibinfo{series}{Leibniz
  International Proceedings in Informatics}, Vol.~\bibinfo{volume}{156})}.
  \bibinfo{publisher}{Schloss Dagstuhl--Leibniz-Zentrum f{\"u}r Informatik},
  \bibinfo{address}{Dagstuhl, Germany}, \bibinfo{pages}{5:1--5:15}.
\newblock
\href{https://doi.org/10.4230/LIPIcs.FORC.2020.5}{doi:\nolinkurl{10.4230/LIPIcs.FORC.2020.5}}


\bibitem[Lou(2025)]%
        {lou2025urbanmas}
\bibfield{author}{\bibinfo{person}{Shangyu Lou}.}
  \bibinfo{year}{2025}\natexlab{}.
\newblock \showarticletitle{Urban-MAS: Human-Centered Urban Prediction with
  {LLM}-Based Multi Agent System}. In \bibinfo{booktitle}{\emph{Proceedings of
  the 3rd ACM SIGSPATIAL International Workshop on Advances in Urban-AI}}.
  \bibinfo{publisher}{Association for Computing Machinery},
  \bibinfo{address}{New York, NY, USA}, \bibinfo{pages}{37--40}.
\newblock
\href{https://doi.org/10.1145/3764926.3771951}{doi:\nolinkurl{10.1145/3764926.3771951}}


\bibitem[Mitchell et~al\mbox{.}(2019)]%
        {mitchell2019modelcards}
\bibfield{author}{\bibinfo{person}{Margaret Mitchell}, \bibinfo{person}{Simone
  Wu}, \bibinfo{person}{Andrew Zaldivar}, \bibinfo{person}{Parker Barnes},
  \bibinfo{person}{Lucy Vasserman}, \bibinfo{person}{Ben Hutchinson},
  \bibinfo{person}{Elena Spitzer}, \bibinfo{person}{Inioluwa~Deborah Raji},
  {and} \bibinfo{person}{Timnit Gebru}.} \bibinfo{year}{2019}\natexlab{}.
\newblock \showarticletitle{Model Cards for Model Reporting}. In
  \bibinfo{booktitle}{\emph{Proceedings of the Conference on Fairness,
  Accountability, and Transparency}}. \bibinfo{publisher}{Association for
  Computing Machinery}, \bibinfo{address}{New York, NY, USA},
  \bibinfo{pages}{220--229}.
\newblock
\href{https://doi.org/10.1145/3287560.3287596}{doi:\nolinkurl{10.1145/3287560.3287596}}


\bibitem[Moreira-Matias et~al\mbox{.}(2015)]%
        {moreiramatias2015porto}
\bibfield{author}{\bibinfo{person}{Luis Moreira-Matias},
  \bibinfo{person}{Michel Ferreira}, {and} \bibinfo{person}{Jo{\~a}o~Mendes
  Moreira}.} \bibinfo{year}{2015}\natexlab{}.
\newblock \bibinfo{title}{Taxi Service Trajectory -- Prediction Challenge,
  {ECML PKDD} 2015}.
\newblock \bibinfo{howpublished}{UCI Machine Learning Repository}.
\newblock
\href{https://doi.org/10.24432/C55W25}{doi:\nolinkurl{10.24432/C55W25}}


\bibitem[Namgung et~al\mbox{.}(2025)]%
        {namgung2025carewell}
\bibfield{author}{\bibinfo{person}{Min Namgung}, \bibinfo{person}{Yao-Yi
  Chiang}, {and} \bibinfo{person}{Olufemi~A. Omitaomu}.}
  \bibinfo{year}{2025}\natexlab{}.
\newblock \showarticletitle{{CareWELL}: Multimodal Region Representation
  Learning with Spatial Contexts for Urban Health}. In
  \bibinfo{booktitle}{\emph{Proceedings of the 3rd ACM SIGSPATIAL International
  Workshop on Advances in Urban-AI}}. \bibinfo{publisher}{Association for
  Computing Machinery}, \bibinfo{address}{New York, NY, USA},
  \bibinfo{pages}{27--36}.
\newblock
\href{https://doi.org/10.1145/3764926.3771947}{doi:\nolinkurl{10.1145/3764926.3771947}}


\bibitem[Pranto et~al\mbox{.}(2025)]%
        {pranto2025hubs}
\bibfield{author}{\bibinfo{person}{Protik~Bose Pranto},
  \bibinfo{person}{Minhazul Islam}, \bibinfo{person}{Ripon~Kumar Saha},
  \bibinfo{person}{Abimelec~Mercado Rivera}, {and} \bibinfo{person}{Namig
  Abbasov}.} \bibinfo{year}{2025}\natexlab{}.
\newblock \showarticletitle{From Hubs to Deserts: Urban Cultural Accessibility
  Patterns with Explainable {AI}}. In \bibinfo{booktitle}{\emph{Proceedings of
  the 3rd ACM SIGSPATIAL International Workshop on Advances in Urban-AI}}.
  \bibinfo{publisher}{Association for Computing Machinery},
  \bibinfo{address}{New York, NY, USA}, \bibinfo{pages}{6--16}.
\newblock
\href{https://doi.org/10.1145/3764926.3771943}{doi:\nolinkurl{10.1145/3764926.3771943}}


\bibitem[Raji et~al\mbox{.}(2020)]%
        {raji2020accountability}
\bibfield{author}{\bibinfo{person}{Inioluwa~Deborah Raji},
  \bibinfo{person}{Andrew Smart}, \bibinfo{person}{Rebecca~N. White},
  \bibinfo{person}{Margaret Mitchell}, \bibinfo{person}{Timnit Gebru},
  \bibinfo{person}{Ben Hutchinson}, \bibinfo{person}{Jamila Smith-Loud},
  \bibinfo{person}{Daniel Theron}, {and} \bibinfo{person}{Parker Barnes}.}
  \bibinfo{year}{2020}\natexlab{}.
\newblock \showarticletitle{Closing the {AI} Accountability Gap: Defining an
  End-to-End Framework for Internal Algorithmic Auditing}. In
  \bibinfo{booktitle}{\emph{Proceedings of the 2020 Conference on Fairness,
  Accountability, and Transparency}}. \bibinfo{publisher}{Association for
  Computing Machinery}, \bibinfo{address}{New York, NY, USA},
  \bibinfo{pages}{33--44}.
\newblock
\href{https://doi.org/10.1145/3351095.3372873}{doi:\nolinkurl{10.1145/3351095.3372873}}


\bibitem[Roberts et~al\mbox{.}(2017)]%
        {roberts2017crossvalidation}
\bibfield{author}{\bibinfo{person}{David~R. Roberts}, \bibinfo{person}{Volker
  Bahn}, \bibinfo{person}{Simone Ciuti}, \bibinfo{person}{Mark~S. Boyce},
  \bibinfo{person}{Jane Elith}, \bibinfo{person}{Gurutzeta Guillera-Arroita},
  \bibinfo{person}{Severin Hauenstein}, \bibinfo{person}{Jos{\'e}~J.
  Lahoz-Monfort}, \bibinfo{person}{Boris Schr{\"o}der},
  \bibinfo{person}{Wilfried Thuiller}, \bibinfo{person}{David~I. Warton},
  \bibinfo{person}{Brendan~A. Wintle}, \bibinfo{person}{Florian Hartig}, {and}
  \bibinfo{person}{Carsten~F. Dormann}.} \bibinfo{year}{2017}\natexlab{}.
\newblock \showarticletitle{Cross-Validation Strategies for Data with Temporal,
  Spatial, Hierarchical, or Phylogenetic Structure}.
\newblock \bibinfo{journal}{\emph{Ecography}} \bibinfo{volume}{40},
  \bibinfo{number}{8} (\bibinfo{year}{2017}), \bibinfo{pages}{913--929}.
\newblock
\href{https://doi.org/10.1111/ecog.02881}{doi:\nolinkurl{10.1111/ecog.02881}}


\bibitem[Scherrer et~al\mbox{.}(2025)]%
        {scherrer2025infrastructure}
\bibfield{author}{\bibinfo{person}{Evan Scherrer}, \bibinfo{person}{Melissa
  Allen-Dumas}, {and} \bibinfo{person}{Bharat Sharma}.}
  \bibinfo{year}{2025}\natexlab{}.
\newblock \showarticletitle{Analyzing Infrastructure Interdependencies Using
  Network-Of-Networks Modeling}. In \bibinfo{booktitle}{\emph{Proceedings of
  the 3rd ACM SIGSPATIAL International Workshop on Advances in Urban-AI}}.
  \bibinfo{publisher}{Association for Computing Machinery},
  \bibinfo{address}{New York, NY, USA}, \bibinfo{pages}{68--71}.
\newblock
\href{https://doi.org/10.1145/3764926.3771946}{doi:\nolinkurl{10.1145/3764926.3771946}}


\bibitem[Yu et~al\mbox{.}(2009)]%
        {yu2009transithub}
\bibfield{author}{\bibinfo{person}{Jie Yu}, \bibinfo{person}{Yue Liu},
  \bibinfo{person}{Gang-Len Chang}, \bibinfo{person}{Wanjing Ma}, {and}
  \bibinfo{person}{Xiaoguang Yang}.} \bibinfo{year}{2009}\natexlab{}.
\newblock \showarticletitle{Cluster-Based Hierarchical Model for Urban Transit
  Hub Location Planning: Formulation, Solution, and Case Study}.
\newblock \bibinfo{journal}{\emph{Transportation Research Record}}
  \bibinfo{volume}{2112}, \bibinfo{number}{1} (\bibinfo{year}{2009}),
  \bibinfo{pages}{8--16}.
\newblock
\href{https://doi.org/10.3141/2112-02}{doi:\nolinkurl{10.3141/2112-02}}


\end{thebibliography}

\end{document}